\documentclass[11pt]{article}

\usepackage{graphics}   
\usepackage{endnotes}            
\usepackage{amsfonts}            
\usepackage{amssymb}             
\usepackage{amsthm}              
\usepackage{amsmath}            
\usepackage{algorithm}           
\usepackage{algorithmic}
\usepackage[letterpaper]{geometry}
\usepackage{rotating}            

\usepackage{setspace}                
\usepackage{natbib}
\bibpunct{(}{)}{;}{a}{,}{;}

\begin{document}

\title{
A Perturbative Solution to the Linear Influence/Network Autocorrelation Model Under Network Dynamics\thanks{The author thanks Weihua An and David Jacho-Chavez for helpful discussions on bias in the network autocorrelation model.}
}

\author{
Carter T. Butts\thanks{Departments of Sociology, Statistics, Computer Science, and EECS, University of California, Irvine; \texttt{buttsc@uci.edu}}
}
\date{10/30/23}
\maketitle

\begin{abstract}
Known by many names and arising in many settings, the forced linear diffusion model is central to the modeling of power and influence within social networks (while also serving as the mechanistic justification for the widely used spatial/network autocorrelation models).  The standard equilibrium solution to the diffusion model depends on strict timescale separation between network dynamics and attribute dynamics, such that the diffusion network can be considered fixed with respect to the diffusion process.  Here, we consider a relaxation of this assumption, in which the network changes only slowly relative to the diffusion dynamics.  In this case, we show that one can obtain a perturbative solution to the diffusion model, which depends on knowledge of past states in only a minimal way.\\[5pt]
\end{abstract}

\theoremstyle{plain}                        
\newtheorem{axiom}{Axiom}
\newtheorem{lemma}{Lemma}
\newtheorem{theorem}{Theorem}
\newtheorem{corollary}{Corollary}

\theoremstyle{definition}                 
\newtheorem{definition}{Definition}
\newtheorem{hypothesis}{Hypothesis}
\newtheorem{conjecture}{Conjecture}
\newtheorem{example}{Example}

\theoremstyle{remark}                    
\newtheorem{remark}{Remark}


The forced linear diffusion model is an important (and sometimes implicit) workhorse of network research, lying behind the feedback centrality scores \citep{koschutzki.et.al:ch:2005}, the Friedkin-Johnsen (F-J) model of social influence \citep{friedkin.johnsen:jms:1990}, and the network (aka ``spatial'') autoregressive model \citep{cliff.ord:bk:1973,anselin:bk:1988,doreian:ch:1989}.  In standard usage, this discrete time model is used to describe the evolution of vertex attributes within a fixed network, where the attribute of each vertex at a given time point is taken to be a linear combination of the attributes of its neighbors at the previous time point, plus an exogenous forcing term.  This model is particularly attractive due to its long-run behavior, which under fairly mild conditions involves convergence to a unique fixed point that depends on the network and the forcing term, but not the past history of the system.  This allows the model to be used in cross-sectional settings (as is typically the case for centrality score and autoregressive model applications), where intertemporal data is not available.  Indeed, this latter use is sufficiently common that many authors using the model make no reference to the underlying dynamic assumptions, and may in some cases be unaware of them.

In real-world settings, however, these assumptions may be consequential.  Among the more obvious is the treatment of the underlying network as fixed; or, more correctly, that any network dynamics are much slower than (i.e., timescale separated from) the convergence of the diffusion process.  The need for timescale separation (and specific justifications for it, in the context of attitudinal influence) is discussed quite explicitly by \citet{friedkin:bk:1998}, but has more often been tacit or unrecognized (e.g., one finds no mention of comparative dynamics in \citet{katz:p:1953,bonacich:ajs:1987,leenders:sn:2002,koschutzki.et.al:ch:2005,butts:ajsp:2008}, among many others).  When timescale separation does not hold, the network is no longer effectively fixed, and the behavior of the linear diffusion model is not in general history-independent.  

Of course, the general notion that power, influence, or other quantities might diffuse on an effectively dynamic network has long been recognized.  Indeed, such joint dynamics are a major motivation for work in the stochastic actor-oriented model (SAOM) tradition \citep{snijders.et.al:ch:2007,steglich.et.al:sm:2010,niezink.snijders:aas:2017}, which has become the dominant approach for studying behavior in this regime.  There has been less interest in the linear diffusion setting, possibly because this breaks the equilibrium condition that makes the model so easily used with cross-sectional data; we review this development in Section~\ref{sec:equilibrium}.  Some exceptions include \citet{friedkin.johnsen:jms:1990,friedkin.johnsen:agp:2003} and work on STARIMA models \citep{pfeifer.deutsch:tibg:1980}; in particular, \citet{friedkin.johnsen:agp:2003} consider a case in which equilibrium is recovered by allowing the network itself to converge to a fixed point, but absent such special cases the interdependence between network and attribute dynamics has traditionally made the model unattractive when time-scale separation is violated.

Here, we return to this classic problem of linear diffusion on a dynamic network without time-scale separation of network and diffusive dynamics.  Although (as we recapitulate) the behavior of this system is in general history dependent, we consider the regime in which network dynamics are relatively slow with respect to the convergence time of the diffusion process.  As we show, this regime allows a perturbative, ``near equilibrium'' solution, in which the attribute distribution can be approximated in terms of the fixed-network equilibrium at the current time point, modified by a correction term that depends only on the local network dynamics (i.e., rates of change).  This solution does not depend on the long-run history of the network, and hence may be more easily used in settings for which only short-term information is available.  The solution also provides insight into how network change impacts the diffusion process.  We demonstrate the solution (and its limitations) with a simulation study of a dynamic social influence process on networks with varying rates of change.

\section{The Basic Model} \label{sec:basic}

The forced linear diffusion model is a discrete-time process defined by the recurrence
\begin{equation}
y_t = A y_{t-1} + z, \label{eq:basemod}
\end{equation}
where $y_t$ is the vector of vertex states at time $t$, $A$ is an $N\times N$ matrix of diffusion weights, and $z$ is an exogenous \emph{forcing vector}, describing the input to each vertex at each time period.  Typically, all quantities are assumed to be real-valued.  Eq.~\ref{eq:basemod} arises in many settings, including several of particular significance in the context of social networks.

\paragraph{Social Influence:} The extensively studied Friedkin-Johnson model of social influence is specified its basic form \citep[e.g. in][]{friedkin:bk:1998} by 
\[
y_t = SG y_{t-1} + (I-S) y^{(0)},
\]
where $S$ is a diagonal matrix whose entries (constrained to the [0,1] interval) are taken to be susceptibilities to influence, $G$ is a matrix of influence weights, and $y^{(0)}$ is usually held to be a vector of initial attitudes (though, as noted below, $y^{(0)}$ will only coincide with an actual time-zero $y_0$ under specific conditions).  $G$ is generally held to satisfy certain conditions, including source conservation and non-negative influence ($0\le G_{ij} \le 1$), convexity ($\sum_{j=1}^N G_{ij}=1$), and the defining condition for self-influence $G_{ii} = 1-S_{ii}$.  Recognizing that we can take $A=SG$ and $z=(I-S) y^{(0)}$, this can be seen as a special case of Eq.~\ref{eq:basemod}.  This further clarifies the role of $y^{(0)}$, which in practice acts as an exogenous source of influence on ego (whose strength is scaled by ego's degree of ``self-influence'').  The identification of $y^{(0)}$ as an initial attitude (i.e., $y_0$) can be rationalized as follows.  Let us assume that each individual is initially in some isolated state, prior to network exposure, in which they are allowed to equilibrate; in this environment, all elements of $S$ and $G$ are identically 0 (since there is no influence), and the vector of attitudes will then converge immediately to $y^{(0)}$.  If the individuals in question are then suddenly put into diffusive contact with each other, they will then start with attitudes $y_0=y^{(0)}$, before evolving due to the influence process.  Although this situation is arguably rare in many naturalistic settings, it accurately mirrors the experimental conditions in which most empirical work on the F-J model has been done, thus motivating the identification of $y^{(0)}$ with the initial attitude in much of the literature.  In general, however, it is more accurate to regard $y^{(0)}$ simply as the amalgam of all influences on network members from outside the network itself, which then diffuse forward as individuals interact.

\paragraph{Feedback Centrality:} Variations of Eq.~\ref{eq:basemod} (or the equilibrium equation described in Section~\ref{sec:equilibrium}) frequently arise in definitions of \emph{feedback centrality} scores \citep{koschutzki.et.al:ch:2005}, often used as \emph{a priori} predictors of power, influence, status, or prestige.  Notable examples include the Katz $\alpha$ index \citep{katz:p:1953}, with
\[
y_t = \alpha W y_{t-1} + \mathbf{1}
\]
where $A=\alpha W$ with scalar parameter $\alpha>0$, and $\mathbf{1}$ is a one-vector.  The Bonacich power score \citep{bonacich:ajs:1987} can similarly be obtained from the recurrence
\[
y_t = \beta W y_{t-1} + \alpha W \mathbf{1}
\]
where $A=\beta W$ with scalar parameter $\beta$ (not necessarily positive), and $\alpha W \mathbf{1}$ for positive scalar $\alpha$ is proportional to the outdegrees of $W$.  A third example is the family of Salancik power indices \citep{salancik:asq:1986}, whose recurrence is given by
\[
y_t = A y_{t-1} + M S
\]
where $M$ is an $n \times m$ matrix of subgroup memberships, and $S$ is an $m$-vector of subgroup ``importance'' scores (with $MS$ taken to be a measure of the ``intrinsic importance'' of each node, apart from the diffusion process).  Perhaps the most famous of all (due to its importance in information retrieval) is PageRank \citep{brin.page:cnIs:1998}, with the recurrence
\[
y_t = \delta P y_{t-1} + (1-\delta) \mathbf{1}
\]
where $\delta \in (0,1)$ is an attenuation parameter, and $P$ is a row-normalized adjacency matrix (in the original context, of hyperlinks).  The similarity to the F-J model is particularly striking, as noted by \citet{friedkin.johnsen:sn:2014}.

As these examples illustrate, most measures of this kind vary primarily by how they define the forcing term (taken as a measure of exogenous power, status, or influence) and whether/how the weight matrix is explicitly scaled (used e.g. in the Bonacich power score to emulate the effects of positive versus negative exchange relationships).  It should be noted that, while not all such scores were originally defined using the recurrent form, the underlying mechanistic logic of an iterative process in which centrality flows through a focal network is ubiquitous in their description. 

\paragraph{Network Autocorrelation:} The network autoregressive (NAR) model \citep[see e.g.][]{cliff.ord:bk:1973,anselin:bk:1988,doreian:ch:1989} is typically specified in terms of the equilibrium solution shown in Section~\ref{sec:equilibrium}, but can be seen as arising from the recurrence
\[
y_t = \alpha W y_{t-1} + X \beta,
\]
where $\alpha$ is a scalar (not necessarily positive), $A = \alpha W$, $X$ is a matrix of nodal covariates, and $\beta$ is a vector of coefficients.  In typical use cases, $W$, $X$, and $y$ are observed (the latter implicitly taken to be an equilibrium realization, with $t \to \infty$), and $\alpha$ and $\beta$ are taken as objects of inference.  The NAR model is better known as the spatial autoregressive model (SAR), due to its widespread use in spatial statistics (where areal units are taken to be nodes, and some measure of contiguity or proximity serves as the weight matrix).  However, there is nothing particularly spatial about the SAR model, and it is in fact simply a rebranded network model.  Because the fit of the diffusion model to real data is never exact, an additional additive error is usually assumed; however, this is added to the equilibrium solution in \emph{ad hoc} fashion, and does not appear in the recurrence.  (In the network moving average (NMA) model, this additive error itself is taken to be the equilibrium of a similar recurrence.  The two may be combined in the network ARMA model, among various other extensions.)  The lack of attention to timescale separation may be particularly concerning in the NAR context, because of its inferential goals: when the system is out of equilibrium, the standard NAR solution will be badly misspecified, leading to biased (and worse, inconsistent) inference.  We comment briefly on this issue in Section~\ref{sec:discussion}.

\subsection{Equilibrium Solution} \label{sec:equilibrium}

In general, the state of the recurrence in Eq.~\ref{eq:basemod} depends on an initial condition $y_0$, i.e.
\begin{gather}
y_1 = A y_0 + z \nonumber \\ 
y_2 = A^2 y_0 + Az + z \nonumber \\
\cdots \nonumber \\
y_t = A^t y_0 + \left[\sum_{i=1}^{t}A^{i-1}\right] z. \label{eq:basetimedep}
\end{gather}
In most applied settings, such a boundary condition is not available, and it may in any event be unclear what timescale is appropriate for selection of the observation time $t$ (i.e., the pace of the updating process may not be known).  However, these issues may in some cases be elided.  In particular, when the spectral radius $\rho(A)$ of the weight matrix is less than unity, $y_t$ will converge to a stable fixed point as $t\to \infty$.  This state is usually derived by positing a hypothetical equilibrium state, $y_\infty$, and noting that it must by definition be a fixed point of the dynamics.  It can then be obtained as
\begin{align}
y_\infty &= A y_\infty + z \nonumber \\
(I-A) y_\infty &= z \nonumber \\
y_\infty &= (I-A)^{-1} z, \label{eq:baseeq}
\end{align}
with the solution existing when $A$ is invertible.  To foreshadow later developments, note that we can also see how this result arises from Eq.~\ref{eq:basetimedep}.  Since $\rho(A)<1$, we have $A^t y_0 \to 0$ as $t \to \infty$ for all finite $y_0$, and the first (``memory'') term containing dependence on $y_0$ vanishes.  The second term in Eq.~\ref{eq:basetimedep} then contains the infinite power series sum $I + A + A^2, \cdots$, which is equal to $(I-A)^{-1}$ under the same spectral radius condition.  Substitution then yields Eq.~\ref{eq:baseeq}.

We may ask what happens when $\rho(A)\ge 1$.  In this regime, the power series will not converge, and typically one observes a divergence of at least some states of $y$; intuitively this is because $A$ is not a long-run contractive mapping, and the associated diffusion process ``blows up'' by iterative amplification.  The conditions of e.g. \citet{friedkin.johnsen:jms:1990} are indeed chosen to avoid this outcome.  (See appendix B of \citet{friedkin:bk:1998} for a particularly extensive discussion of convergence conditions.)  While the system does not converge to a fixed point in this case, it should be noted that the system is still well-defined, and Eq.~\ref{eq:basetimedep} (but not Eq.~\ref{eq:baseeq}) remains valid.  Note that (particularly if $\rho(A)=1$) there may be specific values of $y_0$ and $z$ that do not diverge (an obvious example arising when both are zero-vectors), but stability is guaranteed in general only for $\rho(A)<1$.

The obvious practical value of Eq.~\ref{eq:baseeq} is that it allows us to speak of the behavior of the diffusion process without knowing its initial conditions, and indeed without knowing ``how long'' the system has been running: provided we think that the process of interest has iterated enough to be at (or near) equilibrium, we can describe its behavior using only the weight matrix and the forcing vector.  The plausibility of this interpretation is aided by the fact that convergence to equilibrium is geometric; thus, so long as $A$ is reasonably contractive, few iterations are really needed to be close to equilibrium, and the approximation is easily rationalized.  However, this interpretation still requires a number of assumptions, among them being that $A$ is fixed.  Relaxing this complicates matters, as we now discuss.

\section{Diffusion on Dynamic Networks}

While Eq.~\ref{eq:basemod} presumes a fixed diffusion network, this can be easily relaxed by taking $A$ to be time-dependent.  This leads to the more general recurrence
\begin{equation}
y_t = A_t y_{t-1} + z, \label{eq:timemod}
\end{equation}
where $A_t$ is the matrix of weights active at time $t$.  The consequence of this change can be appreciated by extending the recurrence backward in time to some arbitrary prior period $t-k$:
\begin{align}
y_t &= A_t y_{t-1} + z \nonumber \\
&= A_t [A_{t-1} y_{t-2} + z] + z \nonumber \\
&= A_t A_{t-1} y_{t-2} + [I + A_t] z \nonumber \\
&= A_t A_{t-1} [A_{t-2} y_{t-3} + z] + [I + A_t] z \nonumber \\
&= A_t A_{t-1} A_{t-2} y_{t-3} + [I + A_t + A_t A_{t-1}] z \nonumber \\
&= \left[\prod_{i=0}^{k-1} A_{t-i}\right] y_{t-k} + \left[I + \sum_{i=0}^{k-1} \prod_{j=0}^i A_{t-j}\right] z. \label{eq:timestate}
\end{align}
$y_t$ is now a function of two terms (c.f. Eq.~\ref{eq:timestate}).  The first represents the residual ``memory'' for the state of the system at time $t-k$, while the second represents the accumulated influence of the forcing term through the time-accumulated temporally forward walks through the dynamic graph.  Both, as may be expected, are history dependent: we cannot, in general, know the state of $y_t$ without knowing both an initial condition and the evolution of the diffusion weights in the intervening period.

It is intuitively useful to consider the conventional equilibrium solution in light of Eq.~\ref{eq:timestate}.  Let us imagine that $A$ evolves so slowly that, over the timescale of the attribute dynamics, $A = A_t \approx A_{t-k}$ for all $t,k$.  (I.e., we say that we have \emph{complete timescale separation} between the dynamics of $A$ and the dynamics of $y$.)  In that case, Eq.~\ref{eq:timestate} simplifies to
\begin{align}
y_t &\approx \left[\prod_{i=0}^{k-1} A\right] y_{t-k} + \left[I + \sum_{i=0}^{k-1} \prod_{j=0}^i A\right] z \nonumber \\
&= A^k y_{t-k} + \left[I + A + A^2 + \ldots + A^{k-1}\right] z. \nonumber
\end{align}
This is not by itself sufficient to guarantee equilibrium.  However, if we assume that there exists finite $\delta$ such that $|y_{t-k}| <\delta$ for all $t-k$, and $\rho(A)<1$, then $\lim_{k\to\infty} A^k y_{t-k} = 0$, and we obtain
\begin{align}
y_t &\approx   \left[I + A + A^2 + \ldots \right] z \nonumber\\
&= \left[I-A\right]^{-1} z, \nonumber
\end{align}
where we have exploited the convergence of the matrix power series under our assumed conditions on $A$.  This recapitulates the development of Eq.~\ref{eq:baseeq}, but clarifies what is required for equilibrium to hold in the dynamic matrix case: we need to lose the memory of prior states (which occurs when the matrix product in the first term of Eq.~\ref{eq:timestate} is driven to the zero matrix), and we require convergence of the diffusion of $z$ through the network (which occurs because the high-order terms in the associated matrix sum vanish).  The specific functional form of the solution depends on $A$ being fixed.  Although this last cannot hold in the general case, we observe that the first two conditions do not depend on fixed $A$, and could be generalized to the dynamic setting.  We now turn to this possibility.

\subsection{Behavior with Slow Network Dynamics} 

We now attempt to relax the condition of complete timescale separation, instead invoking a ``slow network dynamics'' condition in which changes in $A$ are weakly dominated by the relaxation in $y$; $y$ is then close to its corresponding static equilibrium, but slightly perturbed by the network dynamics.  Importantly, in this regime we lose long-range dependence on past states, allowing for a solution that can be described without reference to initial conditions.

We begin the development by returning to Eq~\ref{eq:timestate}, and imposing weaker conditions on $A_t$.  We start with the assumption that $A_{t-k}$ is always contractive ($\rho^*(A_{t-k})<1$ for all $t,k$, where $\rho^*$ is the maximum singular value) and that $y$ is bounded (i.e., there exists finite $\delta$ such that $|y_{t-k}|<\delta$ for all $t,k$).  It then follows that  $\lim_{k\to\infty}\left[\prod_{i=0}^{k-1} A_{t-i}\right] y_{t-k} = 0$, leaving us with the limiting expression
\begin{align}
y_t &= \left[I + \sum_{i=0}^{k-1} \prod_{j=0}^i A_{t-j}\right] z \nonumber
\end{align}
which is ``memoryless'' with respect to the past history of $y$.  For convenience in describing dynamics with respect to the focal time $A_t$, we introduce the matrix difference $D_k = A_{t-k}-A_t$, giving us
\begin{align}
y_t &= \left[I + \sum_{i=0}^{k-1} \prod_{j=0}^i (A_t+D_j)\right] z \nonumber\\
&=\left[I + A_t + A_t(A_t + D_1) + A_t(A_t+D_1)(A_t+D_2) \cdots\right] z \nonumber\\
&=\left[I + A_t + A_t^2 + A_t^3 + \cdots +  A_t D_1 + A_t^2 D_1 + A^2_t D_2 + A_tD_1D_2 +  \cdots\right] z \nonumber\\
&=\left[I + A_t\right]^{-1}z + \left[ A_t D_1 + A_t^2 D_1 + A^2_t D_2 + A_tD_1D_2 +  \cdots\right] z. \nonumber
\end{align}
Now, let us consider the slowly varying regime in which $\rho^*(D_k)\ll \rho^*(A_t) < 1$, and $D_k \approx kD$ for small $k$.  We then have
\begin{align}
y_t &\approx \left[I + A_t\right]^{-1}z + \left[ A_t D + A_t^2 D + 2 A^2_t D + 2 A_t D^2 +  \cdots\right] z. \nonumber
\end{align}
Since $D$ and $A$ are ``small'' (i.e., highly contractive), higher powers contribute little to the sum, giving us the first-order dynamic approximation
\begin{align}
y_t &\approx \left[\left[I + A_t\right]^{-1} + A_t D \right] z. \label{eq:timeeq}
\end{align}
This can immediately be seen as a perturbed version of the equilibrium solution in Eq.~\ref{eq:baseeq}, with the solution being linearly displaced from equilibrium by $A_t D z$; intuitively, this is the diffusion of $z$ through all two-walks involving, respectively, the base weight matrix at time $t$ and the change in the weight matrix (which is itself a weight matrix).  As with the past history of $y$, the detailed past history of $A$ is not required: the direction of weight change near the observed time point is sufficient in this regime.  Indeed, it should be observed that the local linearization condition $D_k \approx kD$ is not strictly necessary (although it is convenient): the important condition is $D_k$ is sufficiently contractive that higher-order products involving $D$ contribute little to the final state.

\section{Numerical Examination: Power on a Dynamic Network}

To examine the behavior of the perturbative solution - and, in particular, the extent to which it improves on the fixed equilibrium solution under incomplete timescale separation - we perform a simulation study of power dynamics on evolving organizational networks.  We employ a Salancik-like process \citep{salancik:asq:1986} for power updating, with each individual having an exogenous ``power base'' that acts as a forcing term to their current, realized power level.  I.e., we posit a dynamically evolving ``power score,'' $y_t$, whose evolution is governed by
\[
y_t = \alpha G_t y_{t-1} + f
\]
where $\alpha$ is an attenuation parameter, $G_t$ is the state of the relevant power-transmission network at time $t$, and $f$ is vector of exogenous inputs (representing the basis of each individual's capabilities external to the diffusion process).  

The basis of our model is a case studied by \citet{krackhardt:asq:1990}, who investigated power and influence in a mid-sized information technology company (referred to as ``Silicon Systems'').  Krackhardt reports measurements of perceived friendship and advice-seeking ties within the organization, as well as members' mean ratings of individuals' ``charisma'' (``influence derived from personal magnetism'') and ``potency'' (``ability to get things done that the power-holder wants done'').  In line with the observations of e.g. \citet{krackhardt:asq:1990} and \citet{krackhardt.hanson:hbr:1993} that different relations can be conduits for distinct types of power and influence, we model two hypothetical power-generating processes within the network: ``soft'' power through the friendship network (for which charisma is taken to act as an exogenous source); and power related to expertise or authority through the advice-seeking network (for which potency is taken to act as an exogenous source).  Power in the former is obtained from a combination of charisma and being seen as a friend by charismatic alters (who are themselves seen as friends by charismatic alters), and power in the latter is obtained from a combination of personal potency and being sought-out for advice by potent alters (who are themselves sought-after by potent alters).  While both cases are stylized -- and here employed primarily as an examination of the perturbative solution to the linear updating model under network dynamics -- they thus illustrate basic processes of interest to network researchers.

Network evolution is here modeled using continuous-time ERGM generating processes \citep{butts:jms:2023} whose equilibrium behaviors\footnote{Note that, while the equilibrium of the linear model is a single fixed point, the EGP equilibrium is a graph \emph{distribution}; the process is ergodic, but remains dynamic over time.} are respectively calibrated to the empirically observed friendship and advice networks; as \citet{krackhardt:asq:1990} reports cognitive social structure data (self and proxy reports of ties among all network members), we estimate the underlying network structures using the approach of \citet{butts:sn:2003} (with the graph mixture priors of \citet{butts:tr:2017}, and dividing self and proxy reports as recommended by \citet{lee.butts:sn:2018}).  By varying the rate of change in the ERGM generating process (EGP), we vary the difference between networks at successive time steps, from near-fixed to radically altered.\footnote{Although the EGP itself is continuous, we use it here solely to generate discrete time dynamics that are compatible with linear updating model; this allows us to easily scale the expected changes between time points, but we do not otherwise treat the system as continuous.}  In the former limit, we would expect the fixed-network equilibrium solution to offer a serviceable approximation to the final state of the power dynamics, while in the latter we would expect the system to be too far from equilibrium for a local solution to hold.  Between these extremes, we would expect to see the ``slow dynamics'' regime in which the fixed-network equilibrium fails, but the perturbative solution offers a reasonable approximation to the exact state.  

\subsection{Simulation Study Design}

The simulation study is organized as follows.  We obtain an equilibrium graph distribution by fitting an ERGM to each reference network, then simulating dynamics from an EGP with the fitted ERGM equilibrium.  Using the EGP, we create network time series whose steps reflect differing levels of network change; specifically, we calibrate the EGP such that adjacent networks reflect an average of 1 to 4096 edge change events (in powers of 2).  We generate one time series at each change rate for each of 250 draws from the equilibrium ERGM (using the ERGM draw as the initial state), creating 250 independent replicates at each of 13 change rates.  We simulate time series of 20 network cross-sections, including the initial state.  For each time series, we then simulate a linear updating process, obtaining in each case the final distribution of power scores.  These are then compared with the power scores obtained by a hypothetical equilibrium solution on the final graph in each time series, and with the power scores obtained by the corresponding perturbative solution.  Below, we elaborate on each aspect of the study in detail; results are shown in Section~\ref{sec:results}.

\begin{figure}
\centering
\includegraphics[width=\textwidth]{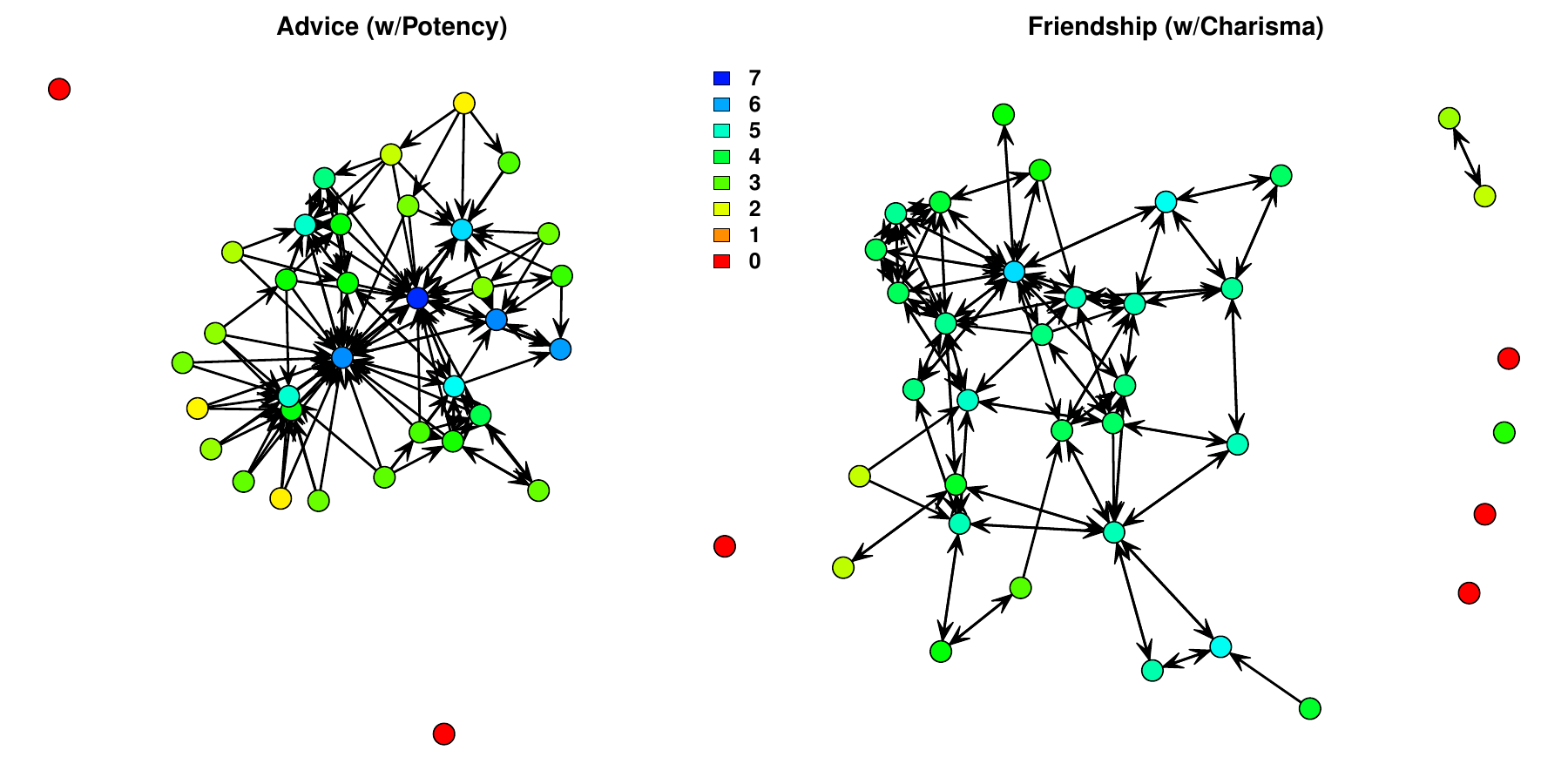}\
\caption{Posterior mode estimates for the advice and friendship networks; nodes are colored by potency and charisma scores (respectively). (Missing edges not shown.) \label{fig:nets}}
\end{figure}

\paragraph{Network Estimation:} We estimate friendship and advice networks for Silicon Systems using the full cognitive social structure (CSS) data, using the protocol of \citet{lee.butts:sn:2018}.  Analysis and data manipulation was performed using the \texttt{sna} \citep{butts:jss:2008b} and \texttt{network} \citep{butts:jss:2008a} libraries from the \texttt{statnet} \citep{handcock.et.al:jss:2008} suite for the \texttt{R} statistical computing system \citep{rteam:sw:2023}.  Symmetric Beta(1,11) error rate priors were used for all informants (with separate self and proxy reporting error rates), and Dirichlet-categorical network priors were used for the graph structure (with hyperparameters set to the Jeffreys prior).  Five independent Markov chains were used for estimation, with 500 burn-in iterations followed by 500 draws per chain.  The edgewise posterior mode was used as the graph estimate for each network.  Because neither self nor proxy reports are available for three actors (and the model hence has no information about their relationships), ties involving these three actors were coded as missing for subsequent ERGM analysis.  Visualizations of the networks are shown in Figure~\ref{fig:nets}.

\paragraph{Equilibrium Network Model:}  To serve as a realistic base case, we use equilibrium distributions of network structures derived from the respective friendship and advice networks.  Specifically, we fit ERGMs to both networks; each network contains ties among the 36 members of the organization, with covariates for organizational rank, assessed potency and charisma, and membership in the potential bargaining unit for a unionization drive occurring during the study period \citep{krackhardt:ch:1992}.  As we interpret being sought for advice or friendship as a source of power to the recipient, we retain the directed character of the data; missing edges are handled using the method of \citet{handcock.gile:aas:2010}.  In addition to baseline density and mutuality effects, we consider differential mixing by membership in the (politically salient) bargaining unit, absolute differences in rank, and target potency and charisma.  (Charisma was neither significant ($p=0.94$) nor AIC-favored in the advice network (AIC 512.4 vs. 479.2), and was thus dropped from this model.)  Absolute difference in rank was favored versus signed difference in rank (even in the advice-seeking network, where the signed effect was neither significant ($p=0.37$) nor AIC-favored (481.3 vs. 497.2)) and hence was used in both models.  Directed GWESP (outgoing two-paths) was employed to capture tendency towards transitive closure.  The models were fit using the \texttt{ergm} library \citep{hunter.et.al:jss:2008,krivitsky.et.al:jss:2023}, using default (Geyer-Thompson-Hummel) settings; adequacy checking was performed using standard simulation-based (``gof'') assessments.  Coefficients for the estimated models are shown in Tables~\ref{tab:adcoef}~and~\ref{tab:frcoef}.

\begin{table}
\centering
\begin{tabular}{lrrrrl}\hline\hline
\multicolumn{1}{c}{Term} & \multicolumn{1}{c}{Estimate} & \multicolumn{1}{c}{Std. Error} & \multicolumn{1}{c}{$z$-value} & \multicolumn{1}{c}{$\Pr(>|z|)$} & \\ \hline 
\texttt{Edges}                          & -7.7357  &   0.6730   & -11.494 & $< 1e-04$ &$***$ \\
\texttt{Mutuals}                         &  0.9529  &   0.4519   &  2.108 & $0.034992$ &$*$ \\ 
\texttt{Nodemix-Bargaining Unit-In$\to$Out}  & -0.5897  &   0.2743   & -2.150 & $0.031567$ &$*$ \\ 
\texttt{Nodemix-Bargaining Unit-Out$\to$In}  & -1.6934  &   0.5817   & -2.911 & $0.003600$ &$**$ \\
\texttt{Nodemix-Bargaining Unit-In$\to$In}   &  0.8659  &   0.2429   &  3.565 & $0.000364$ &$***$  \\
\texttt{Absdiff-Rank}                    & -1.0690  &   0.2035   & -5.252 & $< 1e-04$ &$***$ \\
\texttt{Nodeicov-Potency}                &  1.1633  &   0.1648   &  7.057 & $< 1e-04$ &$***$ \\
\texttt{GWESP-OTP($\phi=0.25$)}            &  1.1509  &   0.2500   &  4.603 & $< 1e-04$ &$***$\\ \hline
\multicolumn{6}{c}{***: $p<0.001$; **: $p<0.01$; *: $p<0.05$; .: $p<0.1$} \\ \hline\hline
\end{tabular}
\caption{\label{tab:adcoef} Parameter estimates for the equilibrium advice network model.}
\end{table}

\begin{table}
\centering
\begin{tabular}{lrrrrl}\hline\hline
\multicolumn{1}{c}{Term} & \multicolumn{1}{c}{Estimate} & \multicolumn{1}{c}{Std. Error} & \multicolumn{1}{c}{$z$-value} & \multicolumn{1}{c}{$\Pr(>|z|)$} & \\ \hline 
\texttt{Edges}                          &-7.950295 &  0.756614  &   -10.508 & $< 1e-04$ &$***$\\
\texttt{Mutuals}                         & 6.556689 &  0.647518  &   10.126 & $< 1e-04$ &$***$\\
\texttt{Nodemix-Bargaining Unit-In$\to$Out} &-0.005268 &  0.485911  &   -0.011 & 0.991351 & \\  
\texttt{Nodemix-Bargaining Unit-Out$\to$In} &-0.676298  &  0.519403  &  -1.302 & 0.192894 &   \\
\texttt{Nodemix-Bargaining Unit-In$\to$In}  & 0.431477 &  0.200935   &  2.147 & 0.031766 & $*$\\  
\texttt{Absdiff-Rank}                   &-0.281091 &  0.119672   &  -2.349 & 0.018832 & $*$\\  
\texttt{Nodeicov-Potency}               & 0.184826 &  0.096936   &  1.907 & 0.056563 & $.$\\  
\texttt{Nodeicov-Charisma}              & 0.611103 &  0.173803   &  3.516 & 0.000438 & $***$\\
\texttt{GWESP-OTP($\phi=0.5$)}           &  0.306050  & 0.103848   &  2.947 & 0.003208 & $**$\\ \hline 
\multicolumn{6}{c}{***: $p<0.001$; **: $p<0.01$; *: $p<0.05$; .: $p<0.1$} \\ \hline\hline
\end{tabular}
\caption{\label{tab:frcoef} Parameter estimates for the equilibrium friendship network model.}
\end{table}

Given the equilibrium network models, we then simulate 250 independent realizations from each to use as starting points for our simulated trajectories.  These were drawn using the standard (Tie/Random Dyad) MCMC algorithm from the \texttt{ergm} library, with a thinning interval of $10^5$.  Note that this process automatically generates edges for all nodes (i.e., since we simulate from the full graph distribution, there are no missing edge states).

\paragraph{Network Evolution Calibration:} We employ longitudinal ERGM (LERGM) dynamics \citep{koskinen.snijders:jspi:2007} for evolution of the underlying networks.  The LERGM EGP for each network is specified by the equilibrium models of Tables~\ref{tab:adcoef}~and~\ref{tab:frcoef}, together with a single rate parameter in each network that governs the overall pace of change.  In the present case, this rate is only meaningful relative to the dynamics of the linear updating model.  To establish a timescale for the network dynamics, we thus adjust the rate so that the expected waiting time to the first event across all 250 equilibrium network draws was approximately 1 (i.e., approximately 1 tie change per unit time, on average).  All hazard calculations and simulations for the EGPs were performed using the \texttt{ergmgp} library \citep{butts:sw:2023}.  

\paragraph{Joint Evolution Model:} Our joint evolution model proceeds as follows.  We begin by generating a sequence of 20 networks starting from each equilibrium network, using realizations from the calibrated EGP initialized with the equilibrium draw; this is done by saving the current network state every $r$ time units, where $r$ is a rate multiplier governing the expected number of tie changes between steps.  (I.e., when $r=16$, adjacent time steps reflect an average of 16 tie change events.)  We then initialize individual power levels with $y_0$ based either on reported potency (for the advice case) or charisma scores (for the friendship case).  Updating is performed using the linear model, with the diffusion matrix $A_t=\alpha G_t$, where $G_t$ is the network state at time $t$, $\alpha = 1/(2\rho_m)$, and $\rho_m$ is the maximum spectral radius observed over the 250 network equilibrium sample (thus ensuring that the model is convergent).  Our focus is then on the comparison of the final power scores at $y_{20}$ from the dynamic model with (1) the state obtained from the fixed-graph equilibrium at $G_{20}$ (i.e., ignoring dynamics), and (2) the state obtained from the perturbative solution at $G_{20}$ (i.e., using the one-step correction).  This is done for the final states arising from each rate multiplier for each initial graph, resulting in a total of $13 \times 250$ sets of comparisons in each case.

\subsection{Results} \label{sec:results}

\paragraph{Equilibrium Models:} Tables~\ref{tab:adcoef}~and~\ref{tab:frcoef} show the coefficients for the ERGM potentials used to drive the respective network dynamics \citep[see][]{butts:jms:2023}.  Both networks show tendencies towards reciprocity (though much more so for friendship than advice-seeking) and transitive closure.  Mixing across ranks is disfavored, and there is a tendency to select alters with higher potency (though this effect is above the 0.05 significance threshold for friendship, its inclusion is favored by AIC/BIC).  Charismatic actors are favored for friendship selection, and there is homophily among members of the bargaining unit; for the advice network, we also see some tendency for unit members to specifically avoid seeking advice from non-members (relative to non-member/non-member ties), and a strong tendency for non-members to avoid seeking advice from unit members.  The LERGM EGP preserves these tendencies, with unfavorable edges breaking more rapidly than favorable ones (and favorable edges forming at a higher rate then ones that would be unfavorable).

\begin{figure}
\centering
\includegraphics[width=\textwidth]{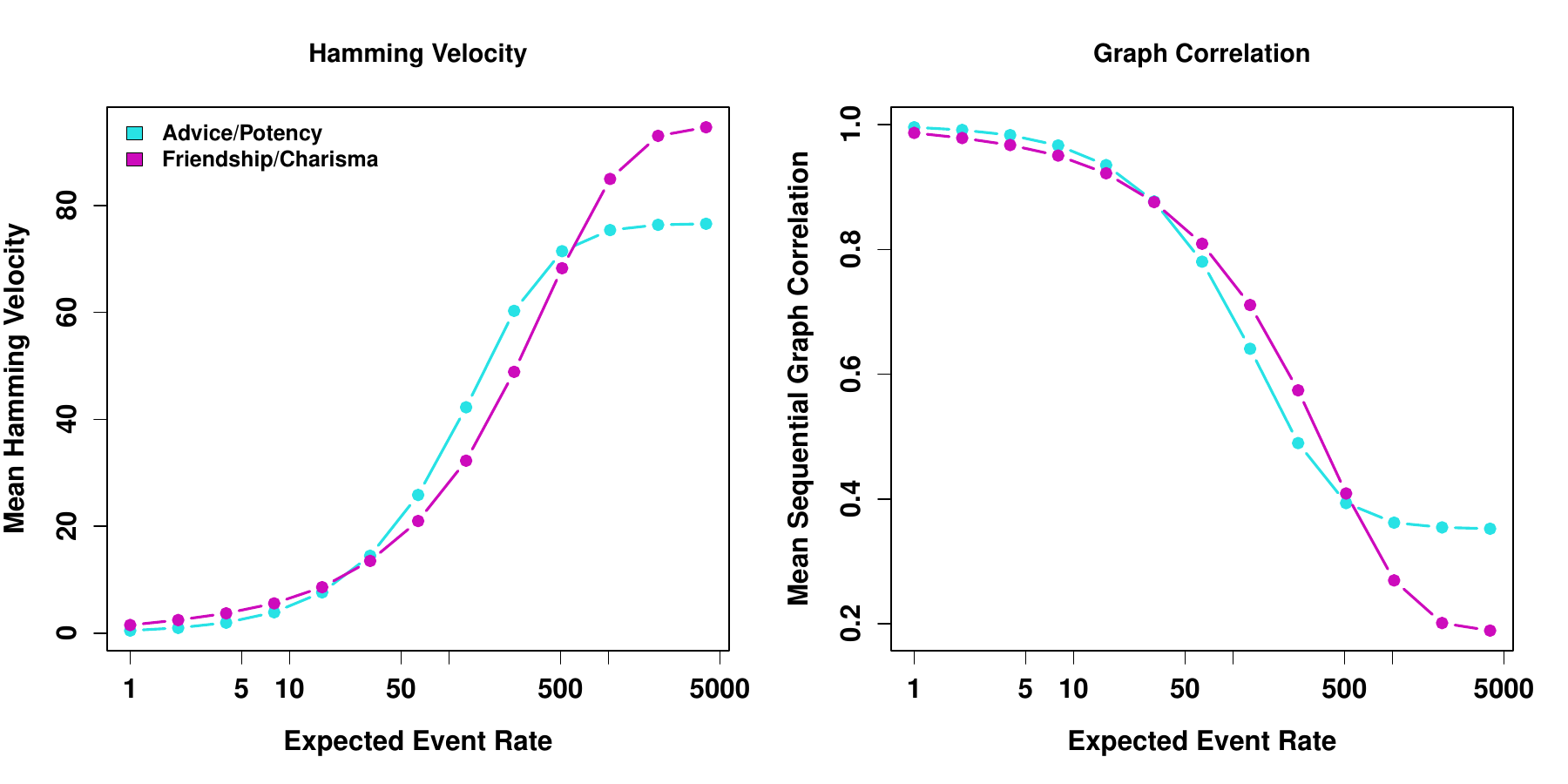}
\caption{Mean change per time step in network structure, as a function of the expected event rate.  Hamming velocity (left) is the number of differences in edge states, while graph correlation (right) is the mean correlation between the respective adjacency matrices.  In both cases, dynamics saturate as one approaches the well-mixed limit. \label{fig:vel}}
\end{figure}

\paragraph{Events Versus the Pace of Network Change:} We simulate trajectories seeded with our 250 equilibrium network draws using an expected $2^0-2^{12}$ tie changes per step; note that tie change events can involve ``churn'' in which the network returns to earlier states, and thus the number of change events is not equivalent to accumulated change per time step.  Figure~\ref{fig:vel} shows the average change between time steps for each model, in terms of Hamming velocity (total edge differences between networks) and graph correlation.  As expected, the realized rate of change increases with the event rate, ultimately leveling off at high event rates.  In this regime, the dynamics are sufficiently well-mixed that one is observing approximately independent draws from the network equilibrium; these networks are still (marginally) correlated, however, due to the presence of covariates that produce persistent structure in the graph.  Figure~\ref{fig:vel} thus confirms that our simulations span the full range from extremely slow dynamics (with roughly one edge change per time step) to ``saturated'' dynamics in which the network states are roughly independent draws from the equilibrium distribution.  Since our interest is in the relationship between approximation error due to ignoring dynamics and the pace of network change, we use the mean Hamming velocity in subsequent analysis.

\begin{figure}
\centering
\includegraphics[width=\textwidth]{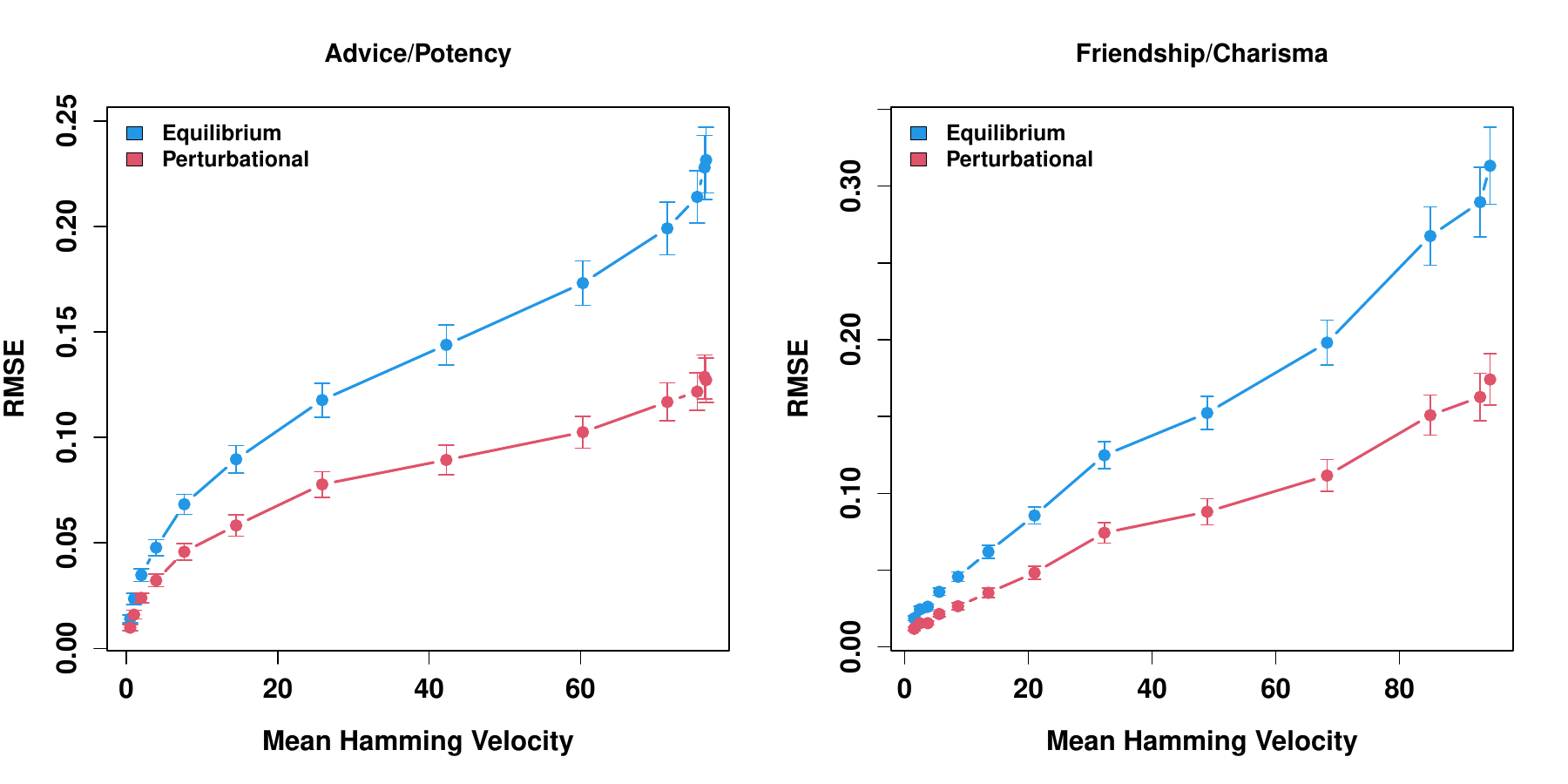}
\caption{Mean approximation error due to use of the fixed-graph equilibrium versus perturbative solutions to final-state power scores, as a function of Hamming velocity.  (Bars indicate 95\% confidence intervals.)  For both systems, the perturbative solution allows comparable error at much higher velocities than the equilibrium solution. \label{fig:acc}}
\end{figure}

\paragraph{Accuracy of the Perturbative Solution:}  Figure~\ref{fig:acc} shows the approximation error (root mean square deviation) for the equilibrium and perturbative power score solutions, versus the actual final states.  For both models (advice/potency and friendship-charisma), approximation errors start near zero in the approximately static case, increasing with the rate of network change; our conditions saturate for very large numbers of edge change events, since (per Figure~\ref{fig:vel}) successive states approach independent draws from the equilibrium ERGM in this regime.  Although errors increase with both approximations, the perturbative solution retains lower levels of error at much higher velocities.  For instance, the error observed for the equilibrium approximation at a rate of 20 edge changes per time step is not attained under the perturbative approximation in the advice network until a rate of approximately 60 edge changes per step; the equivalent gap is 20/40 for equilibrium versus perturbative under the friendship network.  Averaging over the full range of errors produced by both solutions, the perturbative solution obtains a given error level under velocities that are approximately 2.5 times higher for the advice network, and over 4.5 times higher for the friendship network.  It also shows advantages in the high-velocity limit, with the perturbative solution offering a nearly 50\% reduction in error at the extremes.  Thus, while the perturbative solution is still an approximation, it does show consistent improvement on the equilibrium alternative.

\section{Discussion} \label{sec:discussion}

Before concluding, we briefly consider two issues brought into focus by the above results, specifically their implications for inference in the context of network autocorrelation models, and the potential value of fully continuous alternatives.

\subsection{Inferential Implications} 

As noted in Section~\ref{sec:basic}, the difference between the timescale separated equilibrium of the diffusion model and its behavior when timescale separation fails is particularly concerning in network autoregression applications, since the equation used to define the NAR (based on Eq.~\ref{eq:baseeq}) can be a very poor approximation to the behavior of the observed system.  In general, there is no solution that avoids explicit treatment of dynamics; the so-called ``STARIMA'' models \citep{pfeifer.deutsch:tibg:1980} have been proposed for modeling network autocorrelation when all dynamics are observed, but often this is not the case.  In the slow change regime, our result suggests that 
\[
y = \left[\left[I + \rho W\right]^{-1} + \rho W D \right] X \beta + \epsilon,
\]
where $W$ is the observed weight matrix and $D$ is the associated matrix of weight changes, might be a better approximation than the usual (fixed network) model.  This still requires either knowledge of local dynamics (i.e., $D$) or some estimate of thereof, but is less demanding than the fully dynamic solution.  Regardless, we would expect for the conventional solution for $y$ to be off by an additive term approximately equal to $\rho W D X \beta$, which notably involves all model parameters, the weight matrix, the local dynamics, and the attribute covariates.  Inference for both $\rho$ and $\beta$ will thus be impacted by its omission.

\subsection{Continuous Attribute Dynamics} 

To produce the dynamics for our simulation study, we exploited the continuous-time property of the EGP to smoothly vary the rate of change between sequential networks.  The convenience of such continuous-time dynamics strongly suggests applying them to the diffusion model as well.  To date, continuous-time diffusion of continuous attributes has not been widely studied in the social network community, and given the additional complexities involved with such models we have not considered them here.  However, given the quantitative importance of competing timescales for the behavior of diffusion on dynamic networks, moving away from discrete time (which is both cumbersome and arguably unphysical) may have merit.  Indeed, \citet{niezink.snijders:aas:2017} propose a model of exactly this type, albeit with latent network dynamics and stochastic diffusion (a modified Wiener process).  The most natural \emph{deterministic} parallel to the linear diffusion model in the continuous case is the matrix differential equation
\[
\frac{d}{dt} y_t = A y_t + z,
\]
which has the stable fixed point
\[
y_\infty = -A^{-1} z
\]
under appropriate conditions on $A$.  While such models seem of obvious interest for social network applications, a more systematic consideration of the timing of the diffusion process (backed up by more systematic measurement thereof) seems important to effectively constrain them.   

\section{Conclusion} \label{sec:conclusion}

While venerable, the linear diffusion model remains an important tool for modeling processes on graphs.  Given the that much of its appeal comes from its utility in equilibrium modeling, it is useful to ask how far that scenario can be ``stretched'' in the presence of network dynamics.  As shown here, perturbative corrections for out-of-equilibrium behavior are in some cases feasible, leading to a solution that is still independent of long-run history (though dependent on temporally local dynamics).  It is interesting to consider how many other such cases may also be subject to analytical treatment; \citet{friedkin.johnsen:agp:2003}, for instance, consider the case where joint dynamics lead to a stable fixed point for both attitudes and network structure, demonstrating that other solvable regimes exist.  Such developments may further extend the life of this old mathematical workhorse.

\bibliography{ctb}


\end{document}